\documentclass[aps,pre,preprint,superscriptaddress,showpacs,floatfix]{revtex4-1}

\usepackage{epsfig}
\usepackage{latexsym}
\usepackage{amsmath}
\usepackage{amssymb}
\usepackage{amsfonts}
\usepackage{graphicx}
\usepackage{wrapfig}
\usepackage{url}
\usepackage[left]{lineno}

\begin{document}

\title{Experimental proof of Faster-is-Slower in multi-particle systems flowing through bottlenecks}

\author{Jos\'{e} M. Pastor}
\affiliation{Departamento de F\'{i}sica, Facultad de Ciencias, Universidad de Navarra, E-31080 Pamplona, Spain}

\author{Angel Garcimart\'{i}n}
\affiliation{Departamento de F\'{i}sica, Facultad de Ciencias, Universidad de Navarra, E-31080 Pamplona, Spain}

\author{Paula A. Gago}
\affiliation{Departamento de Ingenier\'{i}a Mec\'{a}nica, Facultad Regional La Plata, Universidad Tecnol\'{o}gica Nacional, Av. 60 Esq. 124 S/N, 1900 La Plata, Argentina} 
\affiliation{Consejo Nacional de Investigaciones Cient\'{i}ficas y T\'{e}cnicas, Argentina.}

\author{Juan P. Peralta}
\affiliation{Departamento de Ingenier\'{i}a Mec\'{a}nica, Facultad Regional La Plata, Universidad Tecnol\'{o}gica Nacional, Av. 60 Esq. 124 S/N, 1900 La Plata, Argentina} 

\author{C\'{e}sar Mart\'{i}n-G\'{o}mez}
\affiliation{Departamento de Construcci\'{o}n, Instalaciones y Estructuras, Escuela T\'{e}cnica Superior de Arquitectura, Universidad de Navarra, E-31080 Pamplona, Spain}

\author{Luis M. Ferrer}
\affiliation{Departamento de Patolog\'{i}a Animal, Facultad de Veterinaria, Universidad de Zaragoza, Miguel Servet 177, 50013 Zaragoza, Spain} 

\author{Diego Maza}
\affiliation{Departamento de F\'{i}sica, Facultad de Ciencias, Universidad de Navarra, E-31080 Pamplona, Spain}

\author{Daniel R. Parisi}
\email[]{dparisi@itba.edu.ar}
\affiliation{Instituto Tecnol\'{o}gico de Buenos Aires, 25 de Mayo 444, (1002) C. A. de Buenos Aires, Argentina.}
\affiliation{Consejo Nacional de Investigaciones Cient\'{i}ficas y T\'{e}cnicas, Argentina.}

\author{Luis A. Pugnaloni}
\affiliation{Departamento de Ingenier\'{i}a Mec\'{a}nica, Facultad Regional La Plata, Universidad Tecnol\'{o}gica Nacional, Av. 60 Esq. 124 S/N, 1900 La Plata, Argentina} 
\affiliation{Consejo Nacional de Investigaciones Cient\'{i}ficas y T\'{e}cnicas, Argentina.}

\author{Iker Zuriguel}
\affiliation{Departamento de F\'{i}sica, Facultad de Ciencias, Universidad de Navarra, E-31080 Pamplona, Spain}

\date \today

\begin{abstract}
The ``faster-is-slower'' (FIS) effect was first predicted by computer simulations of the egress of pedestrians through a narrow exit [Helbing D, Farkas I J, Vicsek T, Nature 407:487-490 (2000)]. FIS refers to the finding that, under certain conditions, an excess of the individuals' vigor in the attempt to exit causes a decrease in the flow rate. In general, this effect is identified by the appearance of a minimum when plotting the total evacuation time of a crowd as a function of the pedestrian desired velocity. Here, we experimentally show that the FIS effect indeed occurs in three different systems of discrete particles flowing through a constriction: (a) humans evacuating a room, (b) a herd of sheep entering a barn and (c) grains flowing out a 2D hopper over a vibrated incline. This finding suggests that FIS is a universal phenomenon for active matter passing through a narrowing.
\end{abstract}



\maketitle


Evacuation of humans is a major source of concern during the design of large public facilities, the total evacuation time being the main parameter under consideration. In this respect, computer simulations using the social force model reveal that the evacuation time of a room with a single narrow door depends in a non-trivial way on the preference of pedestrians to exit \cite{Helbing:2000}. In this model the driving force is proportional to the desired velocity of each agent, and simulations show that there exists an optimum desired velocity at which the evacuation time is minimum (and the flow rate is maximum). Further increase of the desired velocity (or driving force) leads to an increase in the evacuation time, a behavior called the faster-is-slower (FIS) effect \cite{Helbing:2000} which is believed to be caused by competitive interaction between the individuals. Indeed, simulations indicate that FIS is associated to the development of an intermittent flow \cite{Helbing:2006}. These intermittencies are caused by the appearance of transient arches that partially clog the outflow \cite{Parisi:2005}, which are possible due to friction forces between contacting individuals \cite{Parisi:2007}.

Although FIS has not been experimentally proved, it is used as validation of the ability of pedestrian models to display realistic behavior \cite{Schadschneider:2002, Kirchner:2003, Hoogendoorn:2004, Wei-Guo:2006, Yamamoto:2007}. Empirical observations of catastrophes seem to endorse the FIS effect since pedestrian clogging and blockage has been observed (for example, The Station nightclub, USA \cite{Fahy:2012, link} and Hillsborough Stadium, England, among others). However, those were a few random observations, and the exact evacuation time dependence on the driving force remains unknown. Besides, controlled tests with humans are difficult due to safety issues and other problems.

Within this context, a natural alternative is to resort to experiments with animals. However, such research has been inconclusive. With scarce exceptions \cite{Saloma:2002}, ants have been used to investigate the passage through bottlenecks. An experiment with escaping ants which were stressed with different repellent concentrations \cite{Soria:2012} revealed a FIS-like curve. However, the increase of the evacuation time at high repellent concentrations was not due to competition or contacts between ants, but to certain damage to the insects' physiology that lead to dysfunction in their motor system \cite{Parisi:2015}. Furthermore, another work shows that ants stressed with temperature present a ``faster-is-faster" response \cite{Boari:2013}. It has been argued that the collective behavior of ants is very particular as they do not interact competitively and do not show phenomena like jamming \cite{Soria:2012,Parisi:2015,Boari:2013,John:2009}.

As discussed in preliminary tests \cite{Garcimartin:2014}, the existence of the FIS effect lacks from a robust and definitive experimental proof. The relevance of filling this gap lies in the fact that validation and deep understanding of the phenomenon will boost our ability to develop new technologies that may prevent undesired long-lasting blockages. It is natural to expect that this phenomenon transcends the case of human throngs and comes about in different instances of particulate materials passing through a bottleneck. The main goals of our work are, first, to provide an experimental proof of the existence of FIS for humans, and second, to demonstrate that it also happens in two other systems ---namely a sheep herd and a granular material--- where discrete elements showing competitive interactions must pass through a constriction.

Usually, the evacuation process is assessed via the evacuation time ($T_N$) for $N$ particles (or individuals), which is defined as the time elapsed between the egress of the first one ($t_1$) and the egress of the $N^{th}$ particle ($t_N$). Here, we will focus on the details of the time lapse $\tau$ between the passage of two consecutive particles ($\tau_i = t_i - t_{i-1}$, for $i \in [2, N] $). These time lapses are related to the evacuation time by $T_N=\sum_{i=2}^{N} \tau_i$. The mean flow rate up to the exit of particle $N$ is then $W = N/T_N = 1 / \langle \tau \rangle$, where the mean lapse time $\langle\tau\rangle$ is taken over the first $N$ particles. A more compact representation is to build, from the $\tau_i$ data, the complementary cumulative distribution function $P(\tau)$ (also know as the reliability or survival function), which gives the probability of finding a time lapse larger than $\tau$. In this representation, the chronology of $\tau_i$ is lost and  $W = 1 / \langle\tau\rangle$ can be calculated only for $N \rightarrow\infty$ by computing $\langle\tau\rangle =\int_0^{\infty}{\tau P(\tau)d\tau}$ whenever the integral exists. Under some conditions, it has been shown that $P(\tau)$ is consistent with the existence of a power-law tail: $P(\tau) \approx \tau^{-\alpha}$. In such cases the exponent $\alpha$ can be used to characterize the dynamics of the egress \cite{Garcimartin:2014,Zuriguel:2014,Gago:2013}.

\begin{figure*}[t]
\begin{center}
\centerline{\includegraphics[width=1\textwidth]{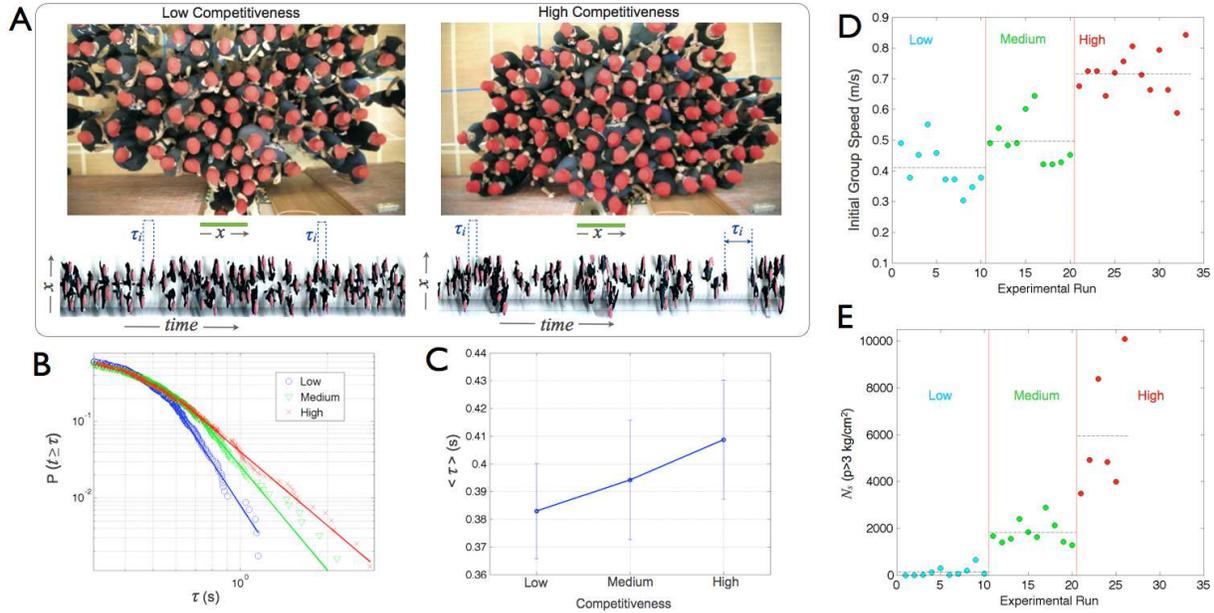}}
\caption{(Color online) Experimental results of the egress of 95 $\pm$ 3 pedestrians through a 69 cm-wide door. (A), The top panels show snapshots of the room during the evacuation drill for a low (left) and high (right) level of competitiveness (i.e., low and high driving force). The lower panels display the corresponding spatio-temporal diagrams. (B), Survival function of the time lapses $\tau$ between two consecutive pedestrians for low (blue circles), medium (green triangles) and high (red crosses) competitiveness. The solid lines represent the fittings obtained with the Clauset-Shalizi-Newman method \cite{Clauset:2009} and the corresponding power-law exponents are $\alpha^{low}~=~6.8~\pm~1$; $\alpha^{medium}~=~5.5~\pm~0.8$ and $\alpha^{high}~=~4.2 \pm~0.4$. (C), Mean $\tau$ as a function of the driving force (i.e., competitiveness level). Error bars indicate 95 \% confidence intervals calculated by bootstrapping. (D-E), Quantification of competitiveness. In (D) maximum group speed when approaching the door at the beginning of the evacuation for each experimental run. Mean speed (dashed lines) and one standard deviation are: $v_d^{low}~=~0.410~\pm~0.071~m/s$; $v_d^{medium}~=~0.497~\pm~0.077~m/s$; $v_d^{high}~=~0.717~\pm~0.075~m/s$. In (E) sum of the number of times that any of the 128 pressure sensors at the doorposts registered a value above $3\ kg/cm^2$. The dashed horizontal lines correspond to the averages obtained for each competitiveness level.}
\label{Fig1}
\end{center}
\end{figure*}

The first experimental system that we consider is a group of pedestrians evacuating a room (Fig. 1). The observations were done at the University of Navarra with 95 volunteers egressing from a room through a 69 cm-wide door. Two video cameras (one inside the room and one outside) registered the evacuation drill from above. From the outside camera, a spatio-temporal diagram is drawn for each evacuation exercise (bottom of Fig. 1A). This graph is built by sampling the same line of pixels from every frame and stacking them. Therefore one axis is the length along the line and the other represents time. In this graph the time lapses ($\tau$) between the passage of two successive pedestrians can be measured.

The volunteers were instructed to evacuate the room as quickly as possible and three levels of competitiveness were studied by asking them to follow these rules: (i) to avoid all physical contact (low competitiveness); (ii) to allow soft contacts with others (intermediate competitiveness), and (iii) to allow soft pushing as a means to make their way out (high competitiveness). We quantify competitiveness by (a) estimating the desired velocity of the pedestrians as the maximum group speed they reach during their initial approach to the door (while the interaction among individuals is very weak, before they crowd in front of the door), and (b) measuring the pressure at one of the doorposts. As shown in Fig. 1D-E, both quantities increase with competitiveness.

In Fig. 1A we display, for low and high competitiveness, a snapshot of the inner camera and the corresponding spatio-temporal diagram obtained from the outer camera. In the high competitiveness scenario, pedestrians arrange in denser configurations in front of the door, and the outflow stream presents a clustered structure. Fig. 1C shows that, as competitiveness increases, so does the mean time lapse $\langle\tau\rangle$. Hence, the flow rate $W$ decreases with increasing driving force, confirming the FIS effect in humans evacuating a room through a narrow door. More information can be obtained by examining the distribution of $\tau$ (Fig. 1B). As recently reported in a pedestrian model and experiments of granular media, colloids and sheep \cite{Zuriguel:2014}, the $\tau$ distributions are compatible with power-law tails; hence, it is convenient to fit the survival function following the Clauset-Shalizi-Newman method \cite{Clauset:2009}. Even though $\langle\tau\rangle$ is only mildly altered for different levels of competitiveness (Fig. 1C), the slopes of the power-law tails display a strong dependence on it. For increasing competitiveness, lower values of the exponent $\alpha$ are obtained, consistent with longer evacuation times. It is worth mentioning that in all these experiments $\alpha > 2$, therefore, $\langle\tau\rangle$  is well defined and so is the mean flow rate $W$.

The second system of self-propelled particles analyzed is a herd of sheep rushing through a door craving for food (Fig. 2). In order to attain different competitiveness, we selected 15 warm days (average temperature $T = 25 \pm 2 ^{\circ}$) and 20 cool days ($T = 10 \pm 5 ^{\circ}$). It is known that sheep activity is affected by temperature \cite{Shinde:2007, Hild:2010} and the desired speed measurements confirmed this hypothesis: lower desired speeds were observed in warm days than in colder days (Fig. 2C). Fig. 2A shows snapshots of the herd at both levels of competitiveness and the corresponding spatio-temporal diagrams used for timing the passage of each animal through the door. As in the case of pedestrians, the survival functions $P(\tau)$ of the time lapses $\tau$ reveal power-law tails whose exponent $\alpha$ decreases as competitiveness increases (Fig. 2D). Undoubtedly,  $\langle\tau\rangle$ increases with the competitiveness level, conclusively proving the FIS effect in a herd of sheep (Fig. 2C).

\begin{figure*}[t]
\begin{center}
\centerline{\includegraphics[width=1\textwidth]{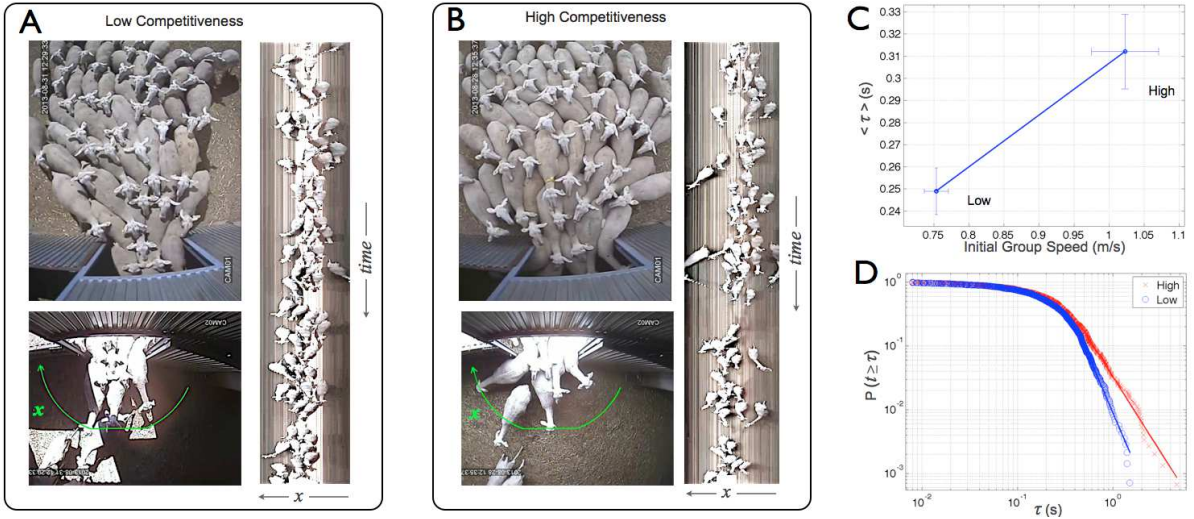}}
\caption{(Color online) Experimental results of $75 \pm 10$ sheep passing through a 94 cm door. (A), Snapshot of the herd for low competitiveness (corresponding to experiments at $T = 25 \pm 2 ^{\circ}$) and (B), high competitiveness (corresponding to experiments at $T = 10 \pm 5 ^{\circ}$) and their associated spatio-temporal diagrams. (C), Average $\tau$ as a function of the initial group speed (which quantifies the competitiveness level). Vertical error bars indicate 95 \% confidence intervals calculated by bootstrapping. Horizontal error bars are standard error. (D), Survival function of the time lapses $\tau$ between the passage of two consecutive sheep for low (blue circles) and high (red crosses) competitiveness. The solid lines represent the fits obtained with the Clauset-Shalizi-Newman method \cite{Clauset:2009}  and the corresponding power-law exponents are $\alpha^{low}~=~4.5~\pm~0.3$; and $\alpha^{high}~=~3.3 \pm~0.2$.}

\label{Fig2}
\end{center}
\end{figure*}

The results presented here are of great relevance as FIS is experimentally demonstrated, providing a quantification of the competitiveness level for the first time. Nevertheless, the intrinsic difficulty of gathering such data in animals and humans prevents a deeper study where the driving force can be fine-tuned. In order to fill this gap, we have resorted to experiments of granular media in a quasi-two-dimensional hopper which is placed over an incline submitted to vibration (Fig. 3A). The angle $\theta$ of the incline allows us to vary the component of the gravity parallel to the hopper plane, granting control on the force that drives the discharge. These systems are known to be prone to clogging \cite{To:2001, Thomas:2015} and vibration is a mechanism to resume the flow \cite{Zuriguel:2014b, Valdes:2008}. The granular material is composed of 500 spherical glass beads and the outlet size is three times wider than the particle diameter.

The survival function for the six angles studied can be seen in Fig. 3B. Given the better temporal resolution attained in this experiment with respect to humans and sheep (where the precision was around $0.1$ s) a distinctive behavior is evidenced for small $\tau$. In particular, the change of curvature at $\tau \approx 0.05$ s suggests that the discharge can be described by the successive alternation of two regimes: (i) a flowing regime when the passage time between successive grains is $\tau<0.05$~s, and (ii) a regime of clogging, for $\tau> 0.05$~s which can be attributed to the formation of arches that are subsequently broken by the vibration. In the regime of clogging we observe a power-law tail (just as in the pedestrian and sheep distributions) also reported in three-dimensional silos \cite{Janda:2009}. Interestingly, the dependence of the survival function on $\theta$ shows opposite trends in the two mentioned regimes: for the flowing part ($\tau<0.05$), increasing $\theta$ reduces $P(t \geq \tau)$; for the clogging regime ($\tau>0.05$), increasing $\theta$ leads to longer $\tau$. This suggests that increasing the driving force leads, at the same time, to an increase in the instantaneous flow rate ($\tau<0.05$) and to an increase of the time needed to break a metastable arch and resume the flow ($\tau>0.05$). This behavior is nicely captured by plotting the mean time lapse for each regime as in Fig. 3D. The flowing regime (left axis) exhibits a faster-is-faster curve similar to a fluid where the flow rate increases monotonically with pressure. On the contrary, the regime of clogging (right axis) shows that the higher the driving force, the smaller the flow rate. The combination of these two trends seems to be at the heart of the characteristic non-monotonicity of the flow rate vs. driving force curves (Fig. 3C). In addition, these results confirm that the origin of the FIS effect is the development of long lasting clogs that lead to a reduction of the mean flow rate.

\begin{figure*}[t]
\begin{center}
\centerline{\includegraphics[width=0.7\textwidth]{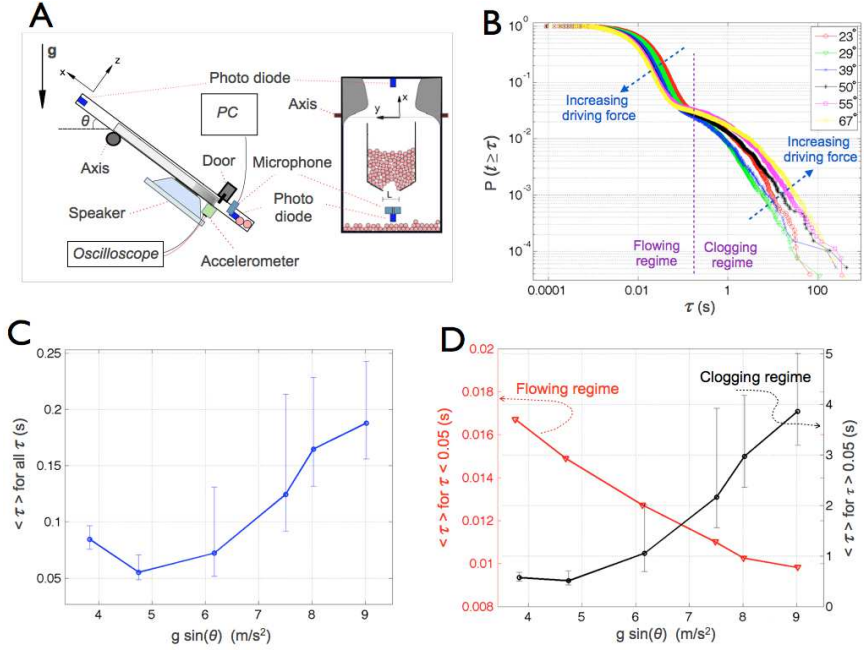}}
\caption{(Color online) Experimental quasi-2D hopper on a vibrated incline. (A), Sketch of the apparatus (side view on the left, and front view on the right). (B), Survival functions of the time lapses $\tau$ between the passage of two consecutive glass beads for six different angles of inclination $\theta$. Arrows indicate the curve trends when increasing the component of the gravity $g \sin (\theta)$ parallel to the incline. (C), Mean $\tau$ as a function of the component of the gravity parallel to the incline. (D), Same results as in Fig. C after splitting the signal in two regimes: for the flowing part we consider only $\tau<0.05$ (red triangles, left axis) and for the clogging episodes only $\tau>0.05$ (black circles, right axis). Note the difference in the scale of both axis. In (C) and (D) error bars or symbol size indicate 95 \% confidence intervals calculated by bootstrapping.}
\label{Fig3}
\end{center}
\end{figure*}

Our findings on these three disparate systems demonstrate the existence of the FIS effect, which can be attributed to the temporary but long-lasting flow arrests caused by intermittent clogging. For the pedestrian dynamics community, these results suppose a benchmark against which the different models should be validated. In particular, we show that increasing competitiveness (which is experimentally characterized by an increase of both desired velocity and pressure against the door) leads to a decrease of the power law exponent of $P(\tau)$. In addition, our observations may support to alternative designs of emergency exits, such as placing an obstacle in front of the door. Even though this contraption implies more complex dynamics than in the experiments reported in this work, it seems reasonable that the pressure reduction accomplished by the obstacle is the cause of the flow improvement \cite{Zuriguel:2014, Helbing:2005, Zuriguel:2011}. Furthermore, observations for sheep and vibrated granular matter suggest that FIS can be a universal phenomenon in active matter and discrete media flowing through narrow constrictions. In this sense, we postulate that the necessary conditions for the observation of the FIS effect are: (a) a discrete flow of particles under a given driving force; (b) a geometrical constriction; (c) a frictional interaction between the particles that allows clog stabilization \cite{Parisi:2007, Suzuno:2013}, and (d) an input of energy (either intrinsic or external) capable of breaking blockades and resuming the flow.

Although potentially any system with these characteristics can display the FIS effect, this has been unnoticed in the literature except in the field of pedestrian dynamics. Future characterization of this effect in different multi-particle systems may boost development in a wide range of applications (such as agribusiness, mineral processing, pharmaceutics, and microfluidics) where the maximization of flow and the minimization of clogging and long-lasting flow arrest of particulate systems are of crucial importance. In particular, in the flow of microparticles through syringes, a phenomenon similar to FIS has been already suggested (increasing the sucking velocity seems to slow down the flow) \cite{Haw:2004}. In the same way, these results provide a well founded explanation to the flow improvement observed when decreasing the fluid velocity of a suspension passing through a porous media \cite{Muecke:1979}, or when reducing the weight of the layer grains in a silo \cite{Zuriguel:2014}. \\

\begin{acknowledgments}
We are indebted to Luis Fernado Urrea for technical assistance with the video systems, Tom\'{a}s Yag\"ue for lending his farm and sheep herd, and all the volunteers that participated in the pedestrian evacuations. We acknowledge University of Navarra Sports Services.\\
This work was funded by: Ministerio de Econom\'{i}a y Competitividad (Spanish Government) through FIS2011-26675 and FIS2014-57325 projects; Mutua Monta\~nesa (Spain); PICT 2011-1238 (ANPCyT, Argentina) and PIUNA (Universidad de Navarra, Spain).
\end{acknowledgments}


\begin{thebibliography}{10}
\bibitem{Helbing:2000} D. Helbing,  I. J. Farkas and T. Vicsek,  Nature \textbf{407}, 487 (2000).
\bibitem{Helbing:2006} D. Helbing, A. Johansson, J. Mathiesen, M. H. Jensen and A. Hansen, Phys. Rev. Lett. \textbf{97}, 168001 (2006).
\bibitem{Parisi:2005} D. R. Parisi and C.O. Dorso, Physica A: Statistical Mechanics and its Applications \textbf{354}, 606 (2005).
\bibitem{Parisi:2007} D. R. Parisi and C.O. Dorso, Physica A: Statistical Mechanics and its Applications \textbf{385}, 343 (2007).
\bibitem{Schadschneider:2002} A. Schadschneider, A. Kirchner and K. Nishinari in {\it Cellular Automata 2002} edited by S. Bandini, B. Chopard and M. Tomassini (Springer, Berlin Heidelberg, 2002), p. 239.
\bibitem{Kirchner:2003} A. Kirchner, K. Nishinari and A. Schadschneider, Phys. Rev. E \textbf{67}, 056122 (2003).
\bibitem{Hoogendoorn:2004} S. P.Hoogendoorn, Transportation Research Record: Journal of the Transportation Research Board, \textbf{1878}, 95 (2004).
\bibitem{Wei-Guo:2006} S. Wei-Guo, Y. Yan-Fei, W. Bing-Hong and F Wei-Cheng, Physica A: Statistical Mechanics and its Applications \textbf{371}, 658 (2006).
\bibitem{Yamamoto:2007} K. Yamamoto, S. Kokubo and K. Nishinari, Physica A: Statistical Mechanics and its Applications \textbf{379}, 654 (2007).
\bibitem{Fahy:2012} R. F. Fahy, G. Proulx and L. Aiman, Fire and materials \textbf{36}, 328 (2012).
\bibitem{link} http://www.youtube.com/watch?v=OOzfq9Egxeo (Accessed 12/10/2014)
\bibitem{Saloma:2002} C. Saloma, G. J. Perez, G. Tapang, M. Lim and C. P. Saloma, Proc. Nat. Acad. Sci. USA \textbf{100}, 11947 (2003).
\bibitem{Soria:2012} S. A. Soria, R. Josens and D. R. Parisi, Safety Science \textbf{50}, 1584 (2012).
\bibitem{Parisi:2015} D. R. Parisi, S. A. Soria and R. Josens, Safety Science \textbf{72}, 274 (2015).
\bibitem{Boari:2013} S. Boari, R. Josens and D. R. Parisi, PloS one \textbf{8}, e81082 (2013).
\bibitem{John:2009} A. John, A. Schadschneider, D. Chowdhury and K. Nishinari, Phys. Rev. Lett. \textbf{102}, 108001 (2009).
\bibitem{Garcimartin:2014} A. Garcimart\'{i}n, I. Zuriguel, J. M. Pastor, C. Mart\'{i}n-G\'{o}mez and D. R. Parisi, Transportation Research Procedia \textbf{2}, 760 (2014).
\bibitem{Zuriguel:2014} I. Zuriguel, D. R. Parisi, R. C. Hidalgo, C. Lozano, A. Janda, P. A. Gago, J. P. Peralta, L. M. Ferrer, L. A. Pugnaloni, E. Cl\'{e}ment, D. Maza, I. Pagonabarraga and A. Garcimart\'{i}n, Scientific Reports \textbf{4}, 7324 (2014).
\bibitem{Gago:2013} P. A. Gago, D. R. Parisi and L. A. Pugnaloni, in {\it Traffic and Granular Flow'11, Moscow, 2011} edited by V. V. Kozlov, A. P. Buslaev, A. S. Bugaev, M V. Yashina, A. Schadschneider and M. Schreckenberg (Springer, Berlin Heidelberg, 2013), p. 317.
\bibitem{Clauset:2009} A. Clauset, C. R. Shalizi and M. E. J. Newman, SIAM Rev. \textbf{51}, 661 (2009).
\bibitem{Shinde:2007} A. K. Shinde and S. A. Karim, Indian Journal of Small Ruminants \textbf{13}, 1 (2007).
\bibitem{Hild:2010} S. Hild, I. L. Andersen and A.J. Zanella, Small Ruminant Research \textbf{90}, 142 (2010).
\bibitem{To:2001} K. To, P. Y. Lai and H. K. Pak, Phys. Rev. Lett. \textbf{86}, 71 (2001).
\bibitem{Thomas:2015} C. C. Thomas and D.J. Durian, Phys. Rev. Lett. \textbf{114}, 178001 (2015).
\bibitem{Zuriguel:2014b} I. Zuriguel, Papers in Physics \textbf{6}, 060014 (2014).
\bibitem{Valdes:2008} J. R. Valdes and J. C. Santamarina, Canadian Geotechnical Journal \textbf{45}, 177 (2008).
\bibitem{Janda:2009} A. Janda, D. Maza, A. Garcimart\'{i}n, E. Kolb, J. Lanuza and E. Cl\'{e}ment, Europhysics Letters \textbf{87}, 24002 (2009).
\bibitem{Helbing:2005} D. Helbing, L. Buzna and A. Johansson, Transportation Science \textbf{39}, 1 (2005).
\bibitem{Zuriguel:2011} I. Zuriguel, A. Janda, A. Garcimart\'{i}n, C. Lozano, R. Ar\'evalo and D. Maza, Phys. Rev. Lett. \textbf{107}, 278001 (2011).
\bibitem{Suzuno:2013} K. Suzuno, A. Tomoeda and D. Ueyama, Phys. Rev. E \textbf{88}, 052813 (2013).
\bibitem{Haw:2004} M. D. Haw, Phys. Rev. Lett. \textbf{92}, 185506 (2004).
\bibitem{Muecke:1979} T. W. Muecke, Journal of Petroleum Technology \textbf{31}, 144 (1979).

\end{thebibliography}
\end{document}